\documentclass[a4paper,10pt]{article}
\usepackage{graphicx,amsmath,natbib}

\def\jcpp{Journal of Chemical  Physics  }

\def\pre{Phys. Rev. E   }
\def\physa{Phys. A    }

\def\za{Z. Astrophys.  } 
\title
{
On non-Poissonian Voronoi tessellations
}
\author{unknown}
\author{M. Ferraro  and L. Zaninetti
\\Dipartimento di Fisica
\\Via Pietro Giuria 1
\\10125, Turin, Italy    \\
\footnote{zaninetti@ph.unito.it}\hspace{0.08cm}
{Corresponding author: zaninetti@ph.unito.it}}
\begin{document}
\maketitle
\begin{abstract}
   The Voronoi
   tessellation is the partition of space
    for a given  seeds  pattern
   and the result of the partition depends completely on
   the type of given pattern
   "random",
 Poisson-Voronoi tessellations
 (PVT),
   or "non-random",
Non Poisson-Voronoi tessellations.
In this note we shall consider properties
of Voronoi tessellations with
centers  generated by
Sobol quasi random sequences which
produce a more ordered disposition of the centers with
respect to the PVT  case.
A probability density function  for volumes of these
Sobol Voronoi tessellations (SVT) will be proposed
and compared with results of numerical simulations.
An application will be presented concerning
the local structure of gas ($CO_2$)
in the liquid-gas coexistence phase.
Furthermore  a probability distribution will be computed
for the length of  chords
resulting from the intersections of
random lines with a  three-dimensional SVT.
The agreement of the analytical formula with the results
from a  computer simulation
will be also investigated.
Finally  a new type of Voronoi tessellation based on
adjustable positions of seeds has been
introduced which generalizes both PVT and SVT cases.
\end  {abstract}
89.75.Da ; Scaling phenomena in complex systems\\

\section{Introduction}
Three-dimensional Voronoi tessellations
produce a random  partition of the space
which have found applications ranging from geology,  \citep{Blower2002}
and molecular biology \citep{Poupon2004,Dupuis2011} to  numerical computing
(for a review see
\citep{Qiang2005}  and references therein),
and chemistry \citep{Jedlovszky2004,Idrissi2011}.

In most studies  Voronoi tessellations have
been considered  in which the
positions of the centers are randomly distributed,
giving rise to the so called
 Poisson-Voronoi tessellations
 (PVT)
\citep{Okabe2000} even though  examples of
non Poissonian Voronoi Tessellation can be found
in the literature \citep{Heinrich1995,Chiu2001,Gonzales2011}.
Non uniform distributions of the centers  can be of interest
to model regular physical configurations;
we shall consider here the properties of Voronoi tessellations
whose center are generated by
Sobol quasi random sequences
\citep{Sobol1967,Bratley1988}.

A probability density function (PDF) for volumes of these
Sobol Voronoi tessellations (SVT) will be proposed
and compared with the results of numerical simulations.

In section \ref{application}  SVT  and  PVT
will be used in an
application concerning the local structure of
gas ($CO_2$) in the liquid-gas coexistence phase.

In addition, we shall consider the relations between
these three-dimensional structures and their
lower dimensional sections;
in particular we shall study chord length distributions resulting
from the random intersection of SVT  with
straight lines.
In case of PVT
probability density functions  of chords can be derived
 rigorously
 \citep{Muche1992,Muche2010},
 see also \citep{Okabe2000};
here we shall present an empirical   method
to calculate a PDF of chord lengths in
the case  of SVT.

Finally Voronoi tessellations derived by perturbating the positions of points
lying on a regular lattice will be considered.

\section{Probability density functions}

\label{pdfsec}

In the case of one-dimensional PVT, with average linear density
$\lambda$, it has been proved   \citep{Kiang1966} that
the distribution of the lengths of the segments has
PDF
\begin{equation}
\label{eq:ki1}
p(l) = 4\lambda^2 l \exp {(-2\lambda l)}.
\end{equation}
At the present time  there are no analytical  formulae for the
area's and volume's distribution and we  limited ourself
to explore the available conjectures.
Numerical experiments
have shown that for PVT an approximate solution can be obtained
via a $3$
parameters generalized
Gamma distribution,
that in case of a variable  $x$ is
\begin{equation}
\label{eq:tanemura}
G(x;a,b,c) =
\frac{
a{b}^{{\frac {c}{a}}}{x}^{c-1}{{\rm e}^{-b{x}^{a}}}
}
{
\Gamma  \left( {\frac {c}{a}} \right)
}
\quad  ,
\end{equation}
see \citep{Hinde1980,Tanemura2003,Khodabin2010,Lazar2013}.
The main moments of  $G(x;a,b,c)$ are:
\begin{equation}
\langle x  \rangle  =
\frac{
{b}^{-{a}^{-1}}\Gamma  \left( {\frac {1+c}{a}} \right)
}
{
\Gamma  \left( {\frac {c}{a}} \right)
}
\quad ,
\end{equation}

\begin{equation}
\sigma^2 =
\frac
{
{b}^{-2\,{a}^{-1}} \left( \Gamma  \left( {\frac {c+2}{a}} \right)
\Gamma  \left( {\frac {c}{a}} \right) - \left( \Gamma  \left( {\frac {
1+c}{a}} \right)  \right) ^{2} \right)
}
{
 \left( \Gamma  \left( {\frac {c}{a}} \right)  \right) ^{2}
 }
 \quad ,
 \end{equation}
 the skewness  $\gamma$ is
 \begin{equation}
 \gamma=
 \frac{
 \Gamma  \left( {\frac {3+c}{a}} \right)  \left( \Gamma  \left( {\frac
{c}{a}} \right)  \right) ^{2}-3\,\Gamma  \left( {\frac {1+c}{a}}
 \right) \Gamma  \left( {\frac {c+2}{a}} \right) \Gamma  \left( {
\frac {c}{a}} \right) +2\, \left( \Gamma  \left( {\frac {1+c}{a}}
 \right)  \right) ^{3}
}
{
 \left( \Gamma  \left( {\frac {c+2}{a}} \right) \Gamma  \left( {\frac
{c}{a}} \right) - \left( \Gamma  \left( {\frac {1+c}{a}} \right)
 \right) ^{2} \right) ^{3/2}
}
\quad ,
\end{equation}
and the  kurtosis $k$
 \begin{equation}
 k= \frac{N_1+N_2}{D}
\quad  ,
\end{equation}
with
\begin{eqnarray}
N_1= & \left( \Gamma  \left( {\frac {c}{a}} \right)  \right) ^{3}\Gamma
 \left( {\frac {4+c}{a}} \right) -4\,\Gamma  \left( {\frac {1+c}{a}}
 \right)  \left( \Gamma  \left( {\frac {c}{a}} \right)  \right) ^{2}
\Gamma  \left( {\frac {3+c}{a}} \right)   \nonumber  \\
N_2 = & 6\, \left( \Gamma  \left( {
\frac {1+c}{a}} \right)  \right) ^{2}\Gamma  \left( {\frac {c}{a}}
 \right) \Gamma  \left( {\frac {c+2}{a}} \right) -3\, \left( \Gamma
 \left( {\frac {1+c}{a}} \right)  \right) ^{4} \nonumber \\
D=& \left( \Gamma  \left( {\frac {c+2}{a}} \right) \Gamma  \left( {\frac
{c}{a}} \right) - \left( \Gamma  \left( {\frac {1+c}{a}} \right)
 \right) ^{2} \right) ^{2}
\quad .
\nonumber
\end{eqnarray}

Usually, instead of $v$ the reduced variable $x=v/\langle v \rangle$
is used \citep{Tanemura2005}:
fitting procedures carried out  in
\citep{Tanemura2005} give  best parameters values
$a=1.16788$, $b= 4.04039$, $c= 4.79803$.

In the next  section we shall adapt the Gamma three-parameter distribution
to fit reduced volumes of Voronoi cells generated with
Sobol sequence.
The results will be compared with those obtained by a simpler PDF,
a one parameter  gamma
\begin{equation}
 p(x;c) = \frac {c^c}{\Gamma (c)}
x^{c-1}
\exp(-cx),
\label{kiang}
\end{equation}
used by Kiang in
his  seminal work on Voronoi tessellations
\citep{Kiang1966}, whose moments are
\begin{equation}
\sigma^2 =\frac{1}{c}, \qquad \gamma=
 2\,{\frac {1}{\sqrt {c}}}, \qquad  k=3\,{\frac {2+c}{c}}
\quad  .
 \end{equation}
It has shown that good approximations for volume distributions of
PVT  cells can be obtained by setting
 $c=5$  \citep{Ferenc_2007}.
A detailed comparison between $p(x;c)$ and $G(x; a,b,c)$ can
be found in
\citep{Ferenc_2007}, see also
\citep{Zaninetti2012f}.

\section{Volumes statistics in SVT}
\label{secnpvt}

Sobol sequences,  like all quasi-random sequences
fill the space more uniformly than
uncorrelated random points and this property
has been extensively used in Monte Carlo methods
such as integration or simulation of transport processes
\citep{Morokoff1994}.
Indeed the uniformity
of quasi-random sequences leads to integration errors  smaller
than in case of random sequences.

\begin{figure}
\begin{center}
\includegraphics[width=10cm]{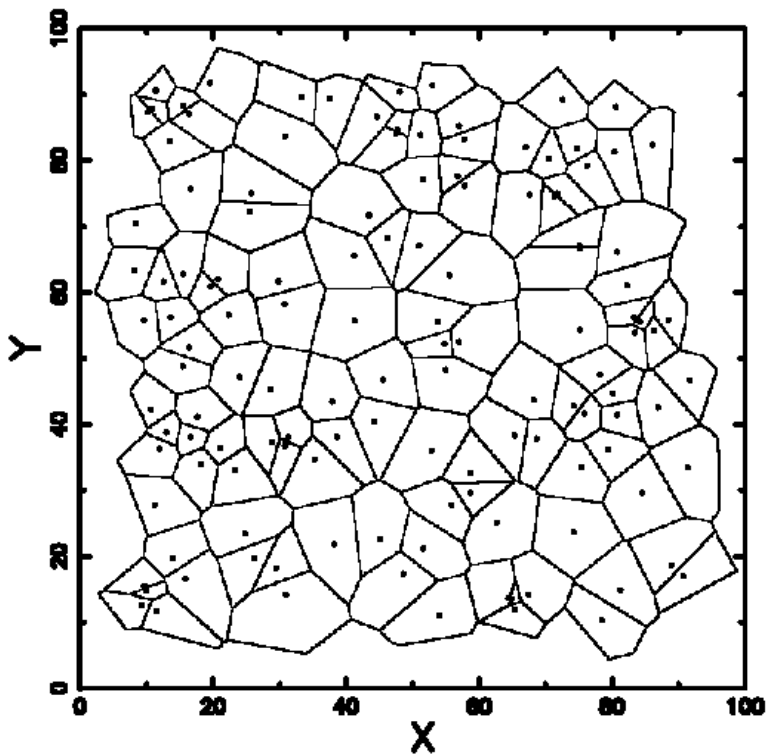}
\end {center}
\caption { An example of PVT  in 2D.
 } \label{area2drandom}
\end{figure}

\begin{figure}
\begin{center}
\includegraphics[width=10cm]{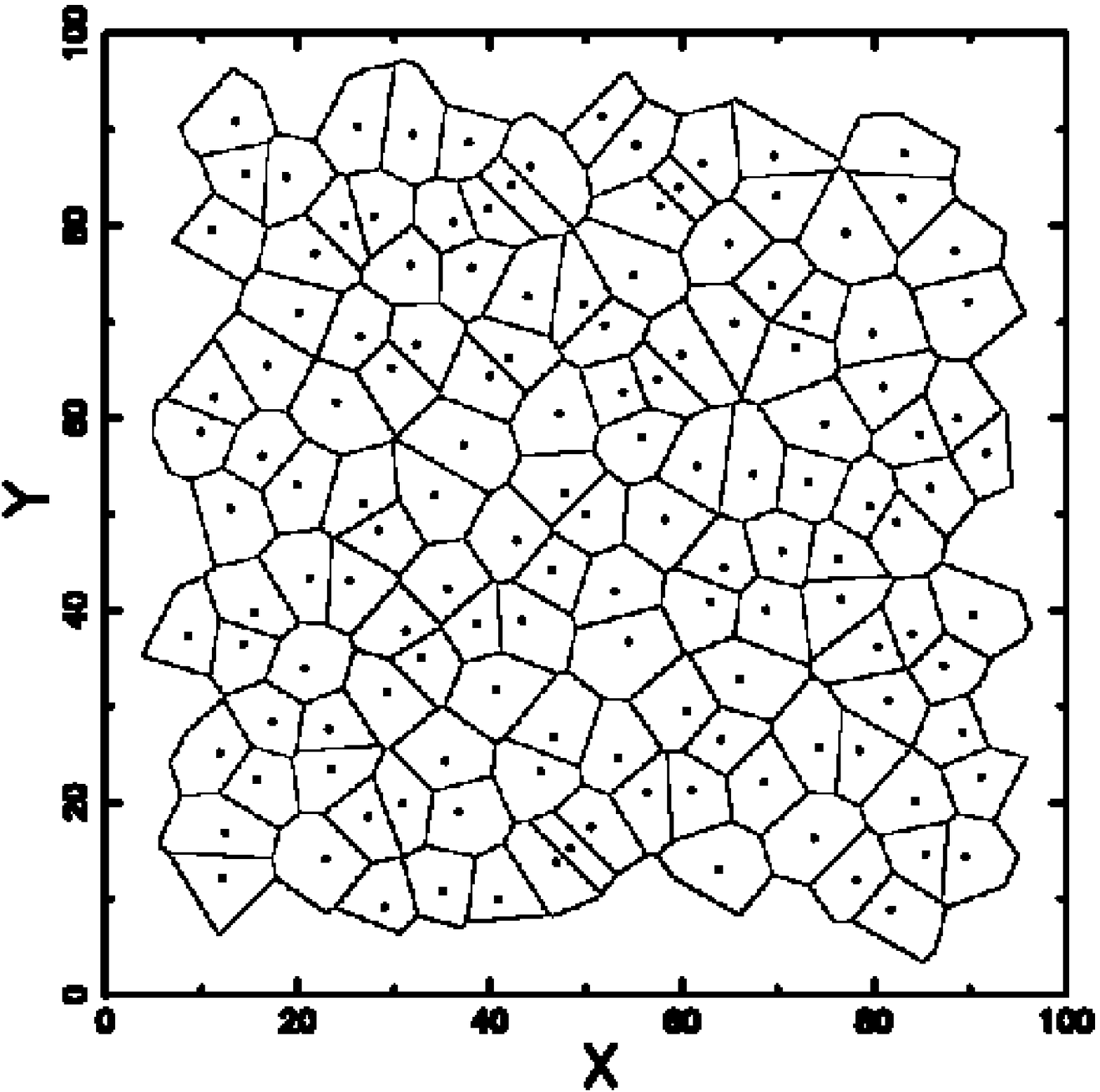}
\end {center}
\caption
{
An example of SVT  in 2D.
} \label{area2dsobol}
\end{figure}

Evidence of uniformity of a Sobol sequence
compared with a random one
is presented in
Figures \ref{area2drandom}
(155 random seeds)
and
\ref{area2dsobol}
(140 quasi-random seeds)
that show respectively  a  PVT and SVT.
Here for the generation of points of the Sobol sequence
 we have used the procedure outlined in \citep{press,Antonov1979},
 for clarity's sake the examples are two-dimensional
and just $140$ centers have been used.
It is apparent from the Figures that SVT exhibits
a narrower distribution of areas: indeed variances are
$0.27$ and $0.061$ for PVT and SVT respectively.

A measure of uniformity  of a quasi-random sequence is the discrepancy,
which is the error made when representing the volume of subsets of the
 unit cube by the fraction of points in the subsets: the lower of the discrepancy
the higher is the uniformity.
There are different ways to define discrepancy
see \citep{Morokoff1994} and references therein,
and it can be shown that the discrepancy on a $d$ dimensional cube
is roughly $(\log n)^sn^{-1}$ for a large number of points $n$, whereas
random sequences have discrepancy of size
$(\log \log n)^{1/2}n^{-1}$.

A simple measure of uniformity  can be also defined as follows:
for  a Sobol tessellation with $n$ seeds of an unit area
 the  minimum distance $d_{min}$
between any two seeds is first computed and  next $d_{min}$ is divided by
 $d_L=\frac{1}{n}$, the distance between two seeds
 in the case of a  regular
lattice of unit area.
Thus the ratio $\frac{d_{min}}{d_L}$ should be larger for a quasi-random sequence of
seeds compared with a random one.
When n=1000 the ratio $\frac{d_{min}}{d_L}$ is
0.093 in the case of 2D Sobol seeds and
0.0056 in the case of random (Poissonian)  seeds.

Thus, Sobol sequences   present a repulsion effect
that can be also found  in
other quasi-random sequences such as, for
example, those generated by  the eigenvalues
of complex random matrices,
see \citep{Lecaer1990}.

\begin{figure}
\begin{center}
\includegraphics[width=10cm]{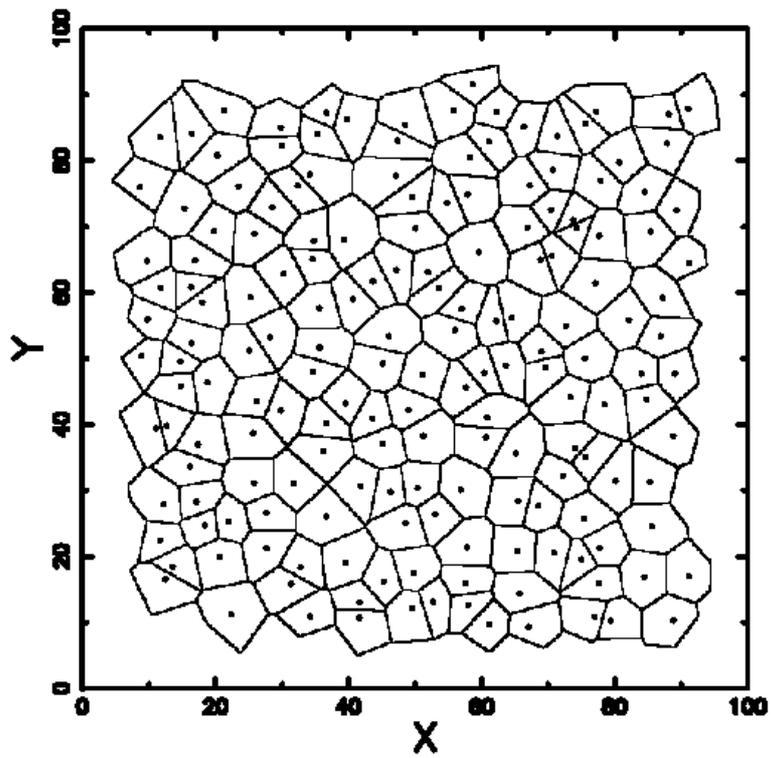}
\end {center}
\caption
{
Tessellation generated
by  174  eigenvalues from complex matrix in 2D.
} \label{area2deigenvalues}
\end{figure}

The eigenvalues  seeds  can be found
starting  from  a random  $N\times N$ complex matrix.
The  matrix elements are given by  $x + iy$ where x and y are pseudo
random real numbers taken from a normal ( Gaussian ) distribution
with mean zero and standard deviation 1/$ \sqrt {2} $.
Once obtained the complex elements we diagonalize
the complex matrix using the subroutine CG from the
EISPACK library.
The points seeds  have the $x$ and $y$ coordinates corresponding
to the real and imaginary parts of the complex eigenvalues
and an example which has variance 0.06
is reported  in Figure
\ref{area2deigenvalues}.

Next we have generated a three-dimensional SVT  with $10^4$
cells: the histogram of their  volumes
has been fitted with
the  generalized gamma $G(x;a,b,c)$, as given by
(\ref{eq:tanemura}), and  the one parameter gamma $p(x;c)$, see
 (\ref{kiang}),  respectively.

The statistical parameters  of the two fits and
the sample's parameters are reported
in Table~\ref{kstest},
were the first line shows numerical values of the parameters
for  generalized three-parameter  and the one-parameter
gamma, respectively, obtained from the fit
of empirical data,
the next lines report values of  mean, variance,
skewness and the kurtosis for the two distributions and the sample,
here the number of cells is $10^4$
and the number of bins is $40$.

\begin{table}[ht!]
\caption
{
Parameters
for  generalized three-parameter  and the one-parameter
gamma, respectively, obtained from the fit
of normalized volumes.
}
\label{kstest}
\begin{center}
\resizebox{12cm}{!}
{

\begin{tabular}{|c|c|c|c|}
\hline
Moment   & Generalized~ gamma  & One~ parameter~ gamma  & Sample  \\
~          & $a=2.3317$, $b= 2.86816$, $c= 7.32528$
&$c= 16.32099$  & ~ \\
\hline
$\langle x_{SVT} \rangle$ &  0.99993          &   1       & 1 \\
\hline
$\sigma^2_{SVT}$&  0.0613809        &   0.06127 & 0.06127  \\
\hline
$\gamma_{SVT}$ &  0.191594         &   0.49505 & 0.20973  \\
\hline
$k_{SVT}$  &  2.94336          &   3.36762 & 3.15243  \\
\hline
\end{tabular}
}
\end{center}
\end{table}

The goodness of the fit has been
assessed by first computing the
PDF
of both $G$ and $p$  and next applying
the Kolmogorov-Smirnov  (K-S) test
\citep{Kolmogoroff1941,Smirnov1948,Massey1951}.

The three- parameter PDF $G$  fits well the simulated volumes:
the maximum distance between the empirical and computed distributions functions
$d_{max}=0.01$ and the result of
the K-S  test, implemented with  the FORTRAN
subroutine KSONE \citep{press},
gives $P_{KS}=0.21$.
Note also that its moments appear
to be close with those derived from the empirical distribution.

\begin{figure}
\begin{center}
\includegraphics[width=10cm]{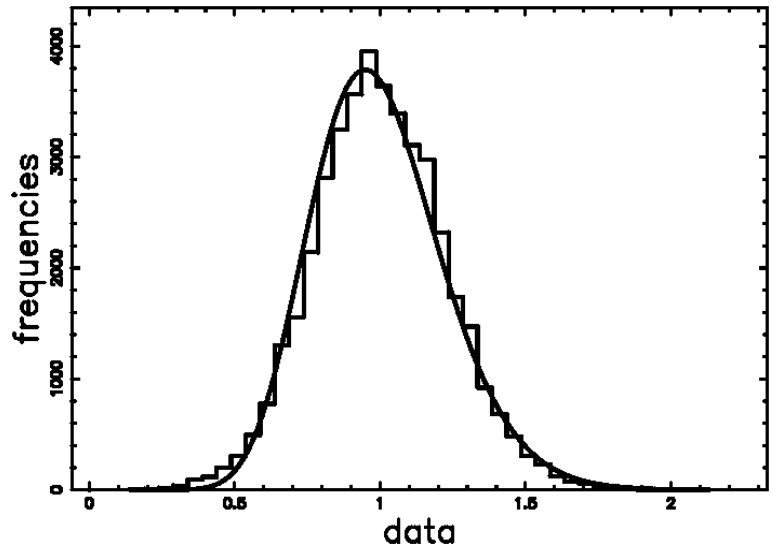}
\end {center}
\caption
{
Histogram (step-diagram) of the SVT
reduced  volume distribution.
}
 \label{volumesobol}%
 \end{figure}

\begin{figure}
\begin{center}
\includegraphics[width=10cm]{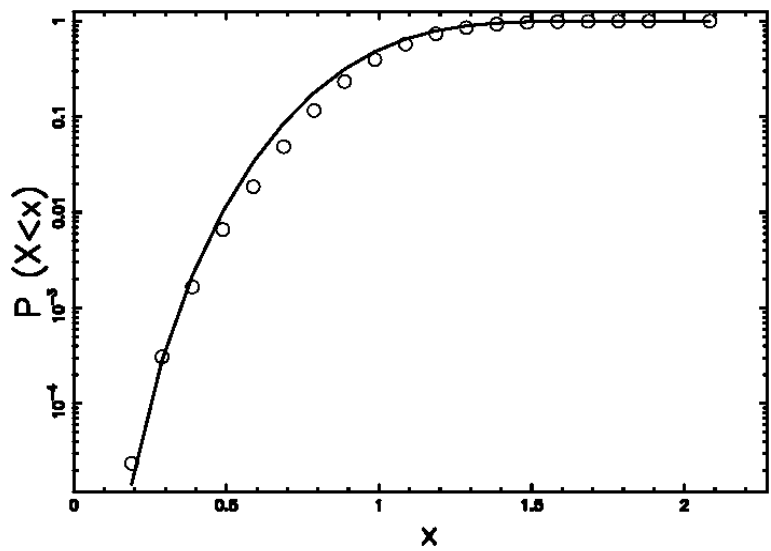}
\end {center}
\caption
{
Simulated  volume distribution (small circles) and
DF of the generalized gamma
(full line).
}
 \label{volumesoboldf}%
 \end{figure}

As concerns  p(x:c)  the
results of the K-S test are worse,
$d_{max}=0.024$, $P_{KS}=1.9\,10^{-5}$.
Figure \ref{volumesobol}
shows the  histogram of
volumes generated by the simulation,
the number of Sobol centers
is $10^4$, the number of division
n=40,
 and the graph of
the generalized gamma used  to fit the data
with parameter values given by Table (\ref{kstest}).
In Figure
\ref{volumesoboldf}
we report
the comparison between
the empirical distribution and
the distribution function (DF) of $G$
with parameters as in Table~\ref{kstest}.

It can be interesting to compare values of the moments obtained here
with those derived for PVT, by using
estimate of $a, b,c$ given in  \citep{Tanemura2005}, namely
$$\sigma^2_{PVT}=0.1787603, \qquad
\gamma_{PVT}= 0.7766972, \qquad
k_{PVT}= 3.849375.$$

Volumes distribution of  SVT  shows a smaller variance, as result of the fact  the centers are distributed in a more regular fashion than in case of PVT;
furthermore the PDF is more symmetric for SVT
$(\gamma_{SVT} < \gamma_{PVT})$;
finally  $k_{SVT}$ is very close to $3$, the  value of  the Gaussian kurtosis.

\begin{figure}
\begin{center}
\includegraphics[width=10cm]{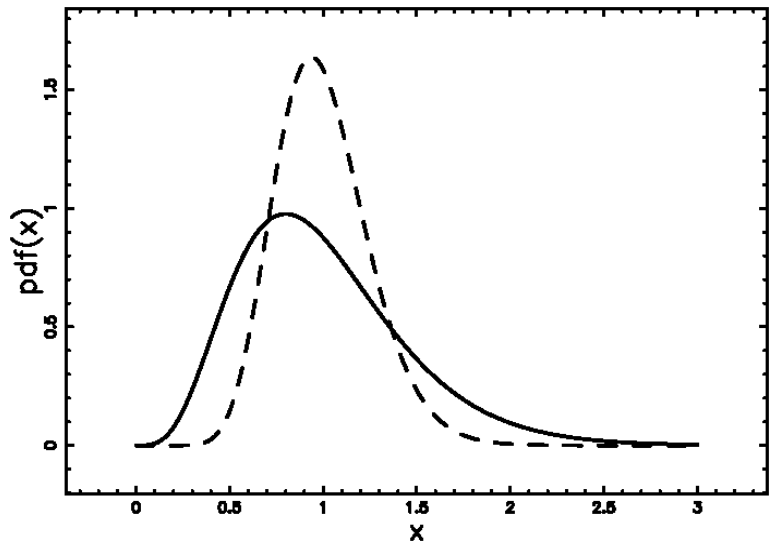}
\end {center}
\caption
{
Plot of reduced gamma PDFs for the PVT
 (full line) and
SVT case (broken line) .
}
 \label{volumesnpvt}%
 \end{figure}

A comparison between the generalized gamma PDFs for the two cases
is shown in Figure  \ref{volumesnpvt}.

\subsection{An application}
\label{application}
In this section  it has been shown that
the main differences between distributions of volumes
in case of SVT vs PVT
are that the former have a smaller variance
and are more symmetric and it has been argued that,
clearly,
 these differences are related to the more
regular distributions of centers in case of
SVTs.
 This suggests a possible application in understanding
different PDF for volumes  obtained in simulations of
local structure of gases,  in the liquid-gas phase.
In \citep{Idrissi2010} $CO_2$ was considered and simulations
were carried out to determine the
volumes available  to each molecule
that were considered as
the center of a Voronoi tessellation.
From the simulation the empirical PDF
of volumes $P(V)$ can be computed
and results show an increase
of mean volume $\langle V \rangle $
and  standard deviation
$\sigma_V$  as temperature rises:
the former effect is due
to the thermal expansion of the system
\citep{Idrissi2010},
whereas the increase of $\sigma_V$ points to
more disordered distribution of the centers
at higher temperatures
and to increasing volume
fluctuations.
To compare our results with the data in \citep{Idrissi2010},
see  Table \ref{tableidrissi},
\begin{table}[ht!]
\caption
{
Mean values and standard deviations of the  Voronoi polyhedra,V,
and c of the Kiang function.
}
\label{tableidrissi}
\begin{center}
\resizebox{12cm}{!}
{
\begin{tabular}{cccccccc}
\hline
 T/K& 250& 270& 285& 298& 303& 306& 313\\
$\langle V \rangle /{\AA}^3$ &  69.7$\pm$  10.3&  77.5$\pm$  13.8&  87.1$\pm$
  19.4&  105.3$\pm$  31.1&  114.4$\pm$  38.2&
  156.8$\pm$  68.8&  156.8$\pm$  65.1\\
 c &  45.79&  31.53&  20.15&  11.46&  8.96&
  5.19&  5.8\\
\hline
\end{tabular}
}
\end{center}
\end{table}
we have computed the standard deviation for the reduced volumes
$x=V/ \langle V \rangle $, namely
$\sigma=\sigma_V/\langle V \rangle$,
and compared them with the  standard
deviations  $\sigma_{SVT}$ and $\sigma_{PVT}$
of the PDFs derived in
\ref{secnpvt}.
As an example when
 $T=250~K$, $\langle V \rangle=69.7{\AA}^3$,
 $\sigma=10.3/69.7=0.147$, $\sigma^2=0.0218$ and $c=1/0.0218=45.79$.
We briefly recall that the data of the previous Table are
theoretical values computed with
the Voronoi polyhedra (VP) analysis based
on numerical codes developed by
\citep{Jedlovszky1999,Jedlovszky2000,Tokita2004}.

In the  temperature range from
$T=250$ to $T=303$  $\sigma$ increases from $0.15$ to $0.33$
and $\sigma_{SVT}=0.25$ is in the middle of this range.
At higher temperatures, $T=303$  and $T=313$ one obtains
$\sigma=0.43$ and $\sigma=0.41 $, respectively,
matching closely
the standard deviation derived for for
PVT,
namely
$\sigma_{PVT}=0.43$.
More importantly, the shape of  the $PDF$,
which is quite symmetric at the lower end of the $T$ range,
increasingly deviates symmetry as $T$ increases and
develops an exponentially decaying tail at high volume values \citep{Idrissi2010}.
This is also what happens in the transition from SVT to PVT,
see Figure   \ref{volumesnpvt}.
These results can be explained as follows:
as $T$ increases positions of molecules become more random
 and a transition takes place from a PDF relatively narrow and  symmetric
(like in the SVT case) to  a more asymmetric
PDF with larger variance corresponding to a PVT.
The relation between the distributions of occupied volumes  and the temperature $T$ can be made clearer
by considering the parameter $c$ of the Kiang distribution (\ref{kiang}).
It is clear that small values of $c$ characterize distributions with
relatively large variance and skweness, whereas as $c$  increase distributions become
narrower and more simmetric.
The parameter $c$ of the Kiang function can be parameterised as
function of the temperature as follows
\begin{equation}
c = C_1 T^{\alpha_1}
\quad  ,
\label{eqnalfa1}
\end{equation}
where $C_1$ and $\alpha_1$ can be found from the data of
Table~\ref{tableidrissi}.
A numerical procedure gives  $C_1= 5.7\,10^{25}$
and $\alpha_1=-10$.

\section{Chords length distribution}
\label{sectionchord}
In many
experimental conditions it is not possible to directly observe
the three-dimensional cells forming a tessellation, just their  linear sections:
thus it is of interest to study the relationships between the
geometric properties of three-dimensional structures and their
chords \citep{Ruan1988,Okabe2000,Stoyan2011}.

One can distinguish
three main ways to generate chords  \citep{Coleman1969}, \citep{Kellerer1984}:
isotropic uniform randomness results when the body is exposed
to an uniform isotropic flow of infinite straight lines; weighted randomness occurs
when a uniformly distributed random point is chosen and is traversed by a
straight line with uniform random direction; two-point randomness is obtained
when
a straight line traverses two random points that are independently and uniformly
distributed.
The first case, which will be considered here since more relevant for practical
applications \citep{Kellerer1984};
relations among PDFs of chords generated by different
methods can be found in  \citep{Kellerer1984}.

\begin{figure}
\begin{center}
\includegraphics[width=10cm]{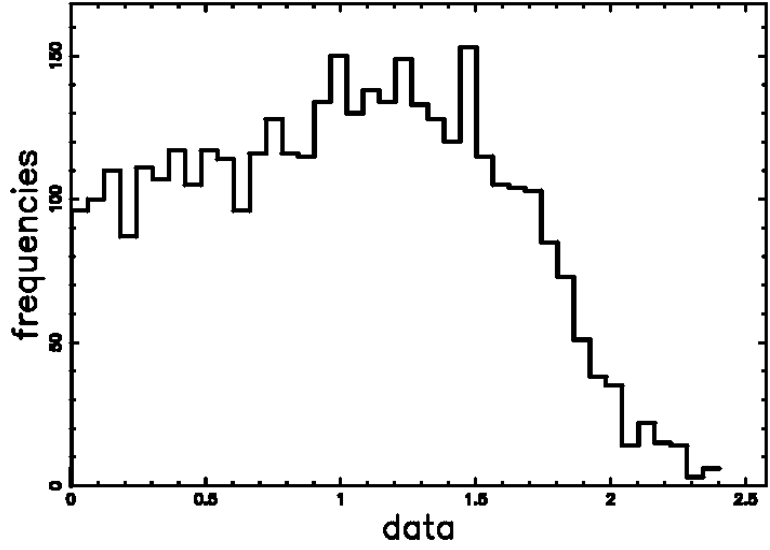}
\end {center}
\caption { Histogram (step-diagram)  for SVT chord
length with
average value $1$ } \label{frequenciessobol}
    \end{figure}

In order to obtain formulas  for the distributions of the chords generated by
the intersections of lines with SVTs
some simplifications  are needed: here  the
polyhedrons forming the cells  will be approximated by
spheres and  the one-parameter distribution $p$ as given  by
Equation(\ref{kiang}), will be used to fit the cells volumes distribution.
With these assumptions a formula for the
PDF of chords length can been obtained,
a simple iterative procedure will then  be used to adapt
this distribution to the simulated data,
thus correcting  the errors resulting from the  approximations.
Let $p_x$ be the  probability density function
for the reduced cell volumes, then
the PDF   $p_y$ for  the lengths of the diameters is given by
\begin{equation}
\label{vtor}
p_y(y)=\frac{\pi}{2} p_x \left (\frac{1}{6}\pi y^3 \right ) y^2
\quad ,
\end{equation}
and the probability density function  $g$  of chords length $l$ is
\begin{equation}
\label{formg}
g(l)=\frac{2l}{\langle y^2 \rangle}\int_l^\infty p_y(y)dy
\quad ,
\end{equation}
see, for instance, \citep{Ruan1988},  \citep{Watson1971}.
In the present case then, with PDF of volumes given  by
the Kiang function, Eq. (\ref{kiang}),
with $c=16$,  the distribution of    diameters is
\begin{equation}
\label{diam}
p_y(y)= \frac{1}{2}\frac{16^{16} \pi ^{16}}{ 6^{15} \Gamma (16)} y^{47}
\exp \left (-\frac{8\pi}{3}  y^3  \right ).
\end{equation}
The use of the generalized gamma function, Eq.(\ref{eq:tanemura}),
for the PDF in volumes means conversely that the integrals which follow
can be done only
in a numerical way.
The probability density  $g_{SVT}$ can be found by
making use of  Eq. (\ref{formg})
 with $p_y$ given by (\ref{diam}), the result is
\begin{equation}
\label{glbnonpoissonian}
g_{SVT}(l)= \frac{a_0}{a_1} \exp \left (- \frac{8}{3} \pi l^3 \right)
 \sum_{k=0}^9 b_k \pi^k l^{3k+1}
\quad,
\end{equation}
where the coefficients are large numbers,
whose values are reported in the Appendix.
The previous formula and the following ones
are  not an exact analytical result but
results from the approximation
of the volume of the
Voronoi's polyhedrons
by spheres.
In order to check the validity of this
PDF
we inserted in a box 50000 seeds
which produce a network of irregular faces
belonging to the Voronoi's  polyhedra.
We selected 120 random lines which will intercept
the network of the irregular faces:
the chord's  length is evaluated as the distance
between a face and the following  one
on the considered line.
A typical run processes a total number of $\approx$ 3200 chords.
The corresponding histogram is shown in
Figure \ref{frequenciessobol}:
note that here the results has been
rescaled so that the average value of chord length  is equal to $1$.
It is apparent that the results of the
simulations do not agree with the PDF given by  Eq.
(\ref{glbnonpoissonian}):  it is enough to note
that $g_{SVT}(0)=0$, is in contrast with the histogram
of Figure   \ref{frequenciessobol}.
In order  to overcome this problem a new  variable  $z$ has
been defined by a shift of $l$: $z=l-a$, so that
$g_{1, SVT}(z)=g_{SVT}(z+a)$. The shift parameter $a$ should not be confused
with the parameter of the generalized gamma PDF (compare Eq.
({\ref{eq:tanemura})).
An explanation for this shift is given by the fact that the length
of the chord which touches only in one point the sphere is zero conversely
a chord which lies on a irregular face of
the Voronoi's polyhedron
has a finite length.
This means that we have more short lengths
in  the simulation of the chords with the real polyhedrons in respect
to the length's of theoretical intersections with the spheres:
i.e. the PDF
of having short chords is finite rather than zero.

 Next, to obtain a reduced variable, a scale change has been
applied resulting in $u =bz\,$.
The scale  parameter $b$ used here  is different from
the parameter of the generalized gamma PDF (compare Eq.
({\ref{eq:tanemura})).
In conclusion,
following translation and   scale change, the final  PDF is now
\begin{eqnarray}
g_{f,SVT}(u;a,b)&=&\frac{C}{b}g_{SVT}\left (\frac{u}{b} +a \right ) \\ \nonumber
&=& \frac{C a_0}{b a_1}  \exp \left (- \frac{8}{3} \pi
 \left (\frac{u}{b}+a \right )^3 \right)
 \sum_{k=0}^9 b_k \pi^k \left ( \frac{u}{b}+a \right )^{3k+1},
\label{glbsobol}
\end{eqnarray}
where $C$ is a normalizing constant,
which has value 1.4717.

Numerical values of $a$ and $b$ have been  obtained by
an interactive procedure that  at each step computes
the DF
\begin{equation}
F(u ;a,b) =\int_0^{u} g_{f,SVT}(z;a,b) dz
\quad :
\end{equation}
 the agreement between the calculated and simulated
distribution function  is then  been verified by
the K-S test.
The procedure
is halted when  $d_{max}$, the maximum  distance between
the distribution functions, reaches a minimum
(that is when the  significance  level  $P_{KS}$ is maximum)
and the corresponding pair $a, b$ is selected;
Table \ref{parameters_seeds} reports the
adopted values.
Calculated and empirical  DFs of chords length
are shown in Figure   \ref{corda_df_sobol},
the K-S test gives $d_{max}=0.0165$, $P_{KS}=0.251$.
The  moments of $g_{f,SVT}$ are presented in
Table \ref{table_parametersnpvt} with
parameters as in Table \ref{parameters_seeds}.
\begin{figure*}
\begin{center}
\includegraphics[width=10cm]{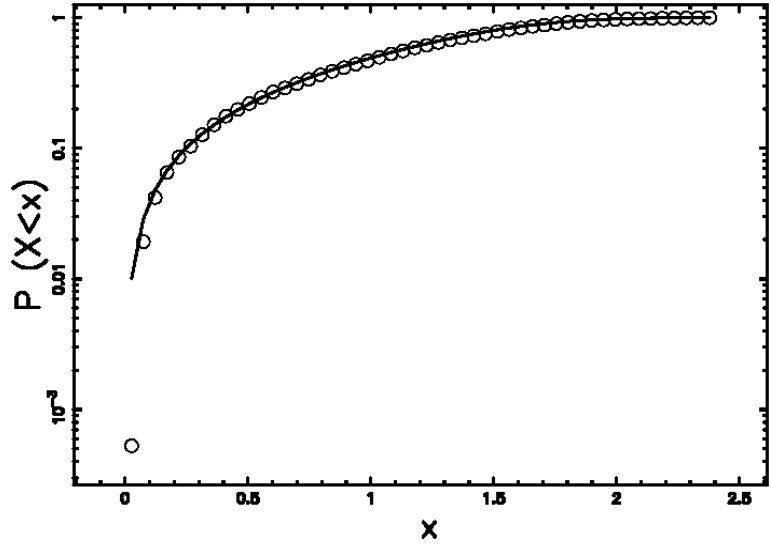}
\end {center}
\caption {
Comparison between data (empty circles) and theoretical
DF for  $g_{f,SVT}$ (continuous line) for chords length.
}
\label{corda_df_sobol}
    \end{figure*}

\begin{table}
 \caption
 {
 Moments of the probability density  function $g_{f,SVT}$, Sobol seeds.
 }
 \label{table_parametersnpvt}
 \[
  \begin{array}{ll}
 \hline
Parameter       & value   \\ \noalign{\smallskip}
 \hline
 \noalign{\smallskip}
Mean            &  1   \\
\noalign{\smallskip} \hline
Variance        & 0.308  \\
\noalign{\smallskip} \hline
Skewness        &  0.0888  \\
 \hline
Kurtosis        &    2.164  \\
 \hline
\end{array}
\]
\end{table}

The results can also be presented as a PDF,
see  Figure   \ref{corda_sobol},
parameters as in Table \ref{parameters_seeds};
the reduced $\chi^2$ is 1.09.
\begin{figure}
\begin{center}
\includegraphics[width=10cm]{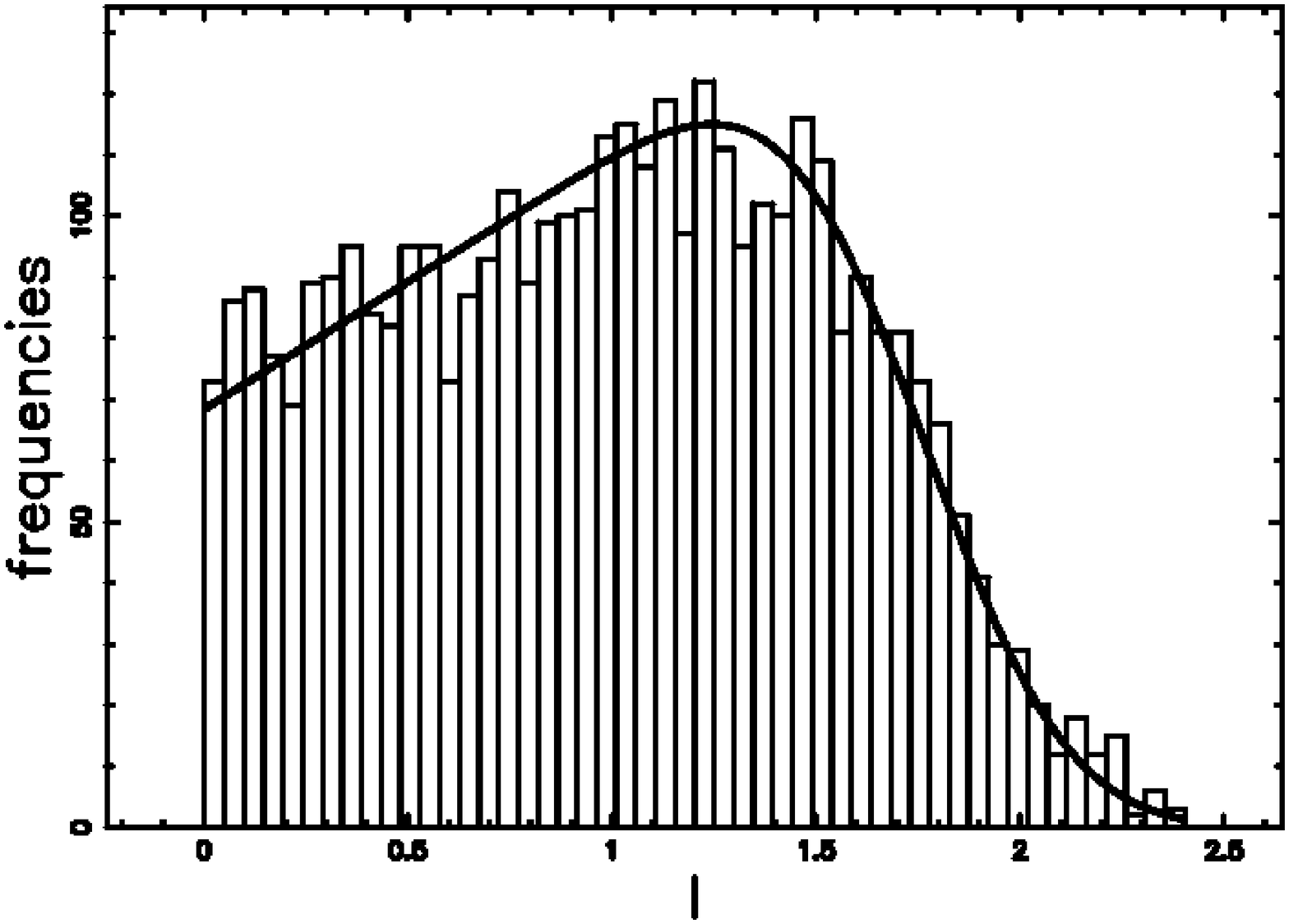}
\end {center}
\caption
{
Histogram (step-diagram)  for SVT chord
length with
average value $1$
and  PDF
$g_{f,SVT}$ (
 line).
 }
\label{corda_sobol}
    \end{figure}
The shifted PDF, $g_{f,SVT}$, for SVT chords can be reported
as a Taylor expansion around x=0 , when the average value is one
\begin{eqnarray}
g_{f,SVT}(x;0.7,3.27)=
0.411+ 0.179\,x- 0.444\,10^{-5}\,{x}^{2}
\nonumber \\
- 0.243 \,10^{-4}\,{x}^{3}- 0.958 \,10^{-3}\,{x}^{4}
 - 0.285 \,10^{-3}\,{x}^{5}  .
\end{eqnarray}
It is clear that at $x=0$ this PDF takes a finite value.

\subsection{The PVT chord}

Chords length distribution in case of PVT can be obtained
with the same method, by adopting now  as the volumes PDF
 Eq. (\ref{kiang}) with $c=5$, which is known to give a
good fit of simulated data \citep{Ferenc_2007}.
The resulting integral for the chord
as given by eqn.(\ref{formg})
is
\begin{eqnarray}
g_{PVT}(l)=
{\frac {125\,{l}^{13}{\pi }^{14/3}{5}^{2/3}\sqrt [3]{6}}{39424\,
\Gamma  \left( 2/3 \right)  \left( {{\rm e}^{\pi \,{l}^{3}}} \right) ^
{5/6}}}+{\frac {75\,{l}^{10}{\pi }^{11/3}{5}^{2/3}\sqrt [3]{6}}{4928\,
\Gamma  \left( 2/3 \right)  \left( {{\rm e}^{\pi \,{l}^{3}}} \right) ^
{5/6}}}
\nonumber  \\
+{\frac {135\,{l}^{7}{\pi }^{8/3}{5}^{2/3}\sqrt [3]{6}}{2464\,
\Gamma  \left( 2/3 \right)  \left( {{\rm e}^{\pi \,{l}^{3}}} \right) ^
{5/6}}}+{\frac {81\,{l}^{4}{\pi }^{5/3}{5}^{2/3}\sqrt [3]{6}}{616\,
\Gamma  \left( 2/3 \right)  \left( {{\rm e}^{\pi \,{l}^{3}}} \right) ^
{5/6}}}
 \\
+{\frac {243\,l{\pi }^{2/3}{5}^{2/3}\sqrt [3]{6}}{1540\,\Gamma
 \left( 2/3 \right)  \left( {{\rm e}^{\pi \,{l}^{3}}} \right) ^{5/6}}}
\nonumber
\quad .
\end{eqnarray}
We now apply  translation and   scale change
\begin{equation}
g_{f,PVT}(u;a,b) =\frac{C}{b}g_{PVT}\left (\frac{u}{b} +a \right )
\quad ,
\label{gfpvt}
\end{equation}
and Table\ref{parameters_seeds} reports the parameters adopted.
\begin{table}[ht!]
\caption {
Parameters which characterize the chord
distribution according to the chosen seed
}
\label{parameters_seeds}
\begin{center}
\begin{tabular}{|c|c|c|c|}
\hline
type ~of ~seed      &  a  &  b   & C \\
\hline
Poissonian~seeds &  0.4 & 1.944413550  & 0.8937531655     \\
Sobol~seeds      &  0.7 & 3.271926163  & 1.471732407      \\
ACG~seeds,s = 0.64 &  0.7 & 3.271926163  & 1.471732407      \\
\hline
\end{tabular}
\end{center}
\end{table}
Figure   \ref{df_pvt}
reports a comparison of the previous
result,
integration of
PDF (\ref{gfpvt})and  parameters as in Table \ref{parameters_seeds},
as a dashed line
 with the tabulated result
as
deduced from Table 5.7.4  in \citep{Okabe2000}
when the average value of both  PDFs is one.
Table \ref{datamuche} reports the two numerical sequences.
\begin{figure*}
\begin{center}
\includegraphics[width=10cm]{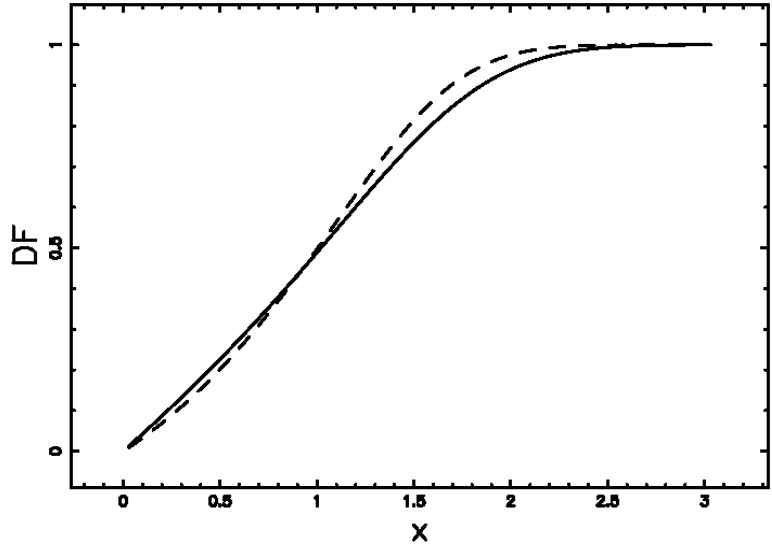}
\end {center}
\caption
{
The full line represents the tabulated
chord  DF  in the case of PVT,
the dashed line is our
chord  DF.
}
\label{df_pvt}
\end{figure*}

\begin{table}
\caption
{
Tabulated  chord  length DF for  PVT case (second column)
and  our DF (third column).
}
 \label{datamuche}
 \[
 \begin{array}{ccc}
 \hline
 \hline
 \noalign{\smallskip}
x    & PVT~DF     & our~DF \\
 \noalign{\smallskip}
 \hline
 \noalign{\smallskip}
0.3 & 0.1320 & 0.1091\\ \hline
0.6 & 0.2760 & 0.2518\\ \hline
0.9 & 0.4336 & 0.4298\\ \hline
1.2 & 0.6008 & 0.6306\\ \hline
1.5 & 0.7602 & 0.8139\\ \hline
1.8 & 0.8844 & 0.9331\\ \hline
2.1 & 0.9579 & 0.9851\\ \hline
2.4 & 0.9891 & 0.9986\\ \hline
2.7 & 0.9981 & 0.9985\\ \hline
3.0 & 0.9998 & 0.9983\\ \hline
 \hline
 \end{array}
 \]
 \end{table}

Calculated and empirical  DFs of chords length
in the PVT case are shown in Figure
\ref{corda_df_poisson} with
parameters as in Table \ref{parameters_seeds};
the K-S test gives $d_{max}=0.0362$.
\begin{figure*}
\begin{center}
\includegraphics[width=10cm]{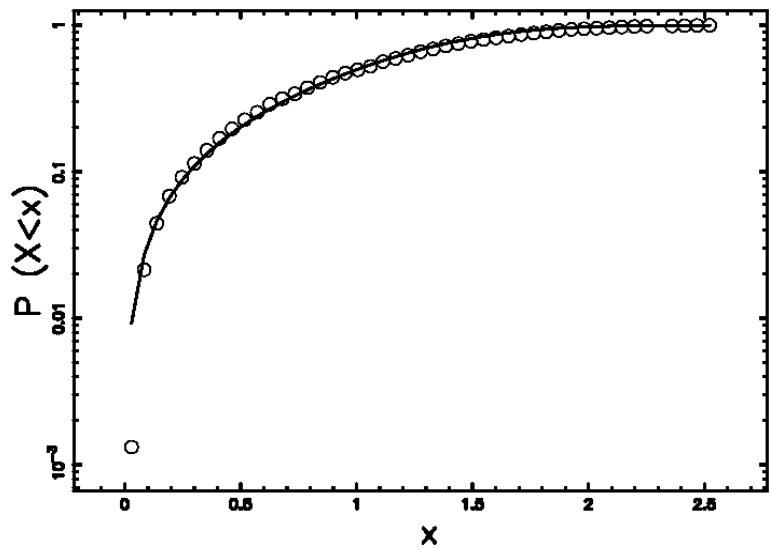}
\end {center}
\caption {
Comparison between data (empty circles) and theoretical
DF for   $g_{f,PVT}$ (continuous line)
of chords length distribution.
}
\label{corda_df_poisson}
    \end{figure*}

For comparison purposes a plot of PDF
$ g_{f}(u;a,b) $, which represents the SVT case with
parameters as in Table \ref{parameters_seeds},
is shown in Figure
\ref{cordavsd}
together with the numerical PDF for
PVT derived from the numerical DF
reported in Table 5.7.4 of \citep{Okabe2000};
in both cases the mean chord length is equal to $1$.
\begin{figure*}
\begin{center}
\includegraphics[width=10cm]{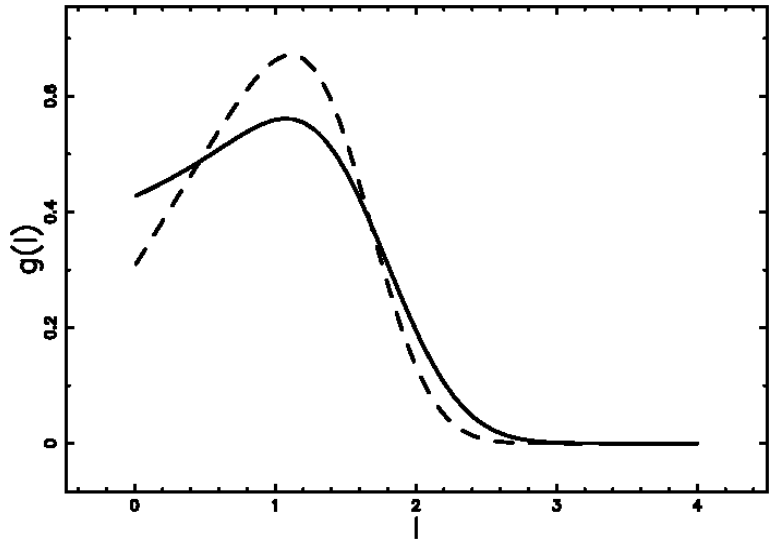}
\end {center}
\caption
{
The full line represents the PDF of chords length in case of PVT,
the dashed line is the graph of $g_{f,SVT}$.
}
\label{cordavsd}
\end{figure*}
The shifted PDF, $g_{f,PVT}$, for PVT chords is reported
as a Taylor expansion around $x=0$, when the average value is one
\begin{eqnarray}
g_{f,PVT}(x;0.563,2.452 )=
0.386+ 0.279\,x- 0.00257\,{x}^{2}\nonumber \\
-  0.00766\,{x}^{3}- 0.0151\,{x}^{4}- 0.02\,{x}^{
5}
\, .
\end{eqnarray}

\section{Adjustable seeds}

An occasional reader may question
if a   scenario of  gradual  transition
from PVT to SVT can be outlined.
In order to have more  flexible  seeds we   introduce
the  adjustable  Cartesian grid  (ACG)
which  can be computed both in 2D and 3D.

The algorithm is now outlined:
\begin{enumerate}
\item  The process starts   inserting the seeds
       on a 2D/3D regular
       Cartesian grid with equal distance $\delta$ between
       one point and  the following one
\item  A random radius is generated according to the
       half Gaussian ,$HN(x)$, which  is defined
       in the interval $[0,\infty]$
\begin{equation}
HN(x;s) =
\frac {2} {s (2 \pi)^{1/2}} \exp ({- {\frac {x^2}{2s^2}}} )
\quad
0 < x < \infty
\quad .
\label{gaussianhalf}
\end{equation}
The main moments of  $HN(x;s)$ are:
\begin{equation}
\langle x \rangle  =
{\frac {s\,\sqrt {2}}{\sqrt {\pi }}}
\quad ,
\end{equation}
and
\begin{equation}
\sigma^2 = {\frac {{s}^{2} \left( \pi -2 \right) }{\pi }}
\quad .
 \end{equation}
\item
A random direction is chosen in 2D/3D and the two/three
Cartesian coordinates  of  the generated radius
are evaluated.
These two/three   small Cartesian components
are  added to the regular 2D/3D grid  which represent
the seeds.
In order to have  small corrections we express
$s$  in $\delta$ units.
The parameter $s$  is a
 good "disorder parameter" for
the generated configurations.
At  $s=0$  we will have the seeds disposed on  a  perfect lattice
with all the volumes of the irregular polyhedra equal,
increasing s we will reach  before c=16  in the PDF of volumes (SVD)
and  subsequently c=5  (PVT).
\end{enumerate}
Figure
\ref{area2dlattice2v}
reports an example
of 2D tessellation from ACG  which
areas have variance 0.047, the same value of
the 2D sobol seeds.
\begin{figure}
\begin{center}
\includegraphics[width=10cm]{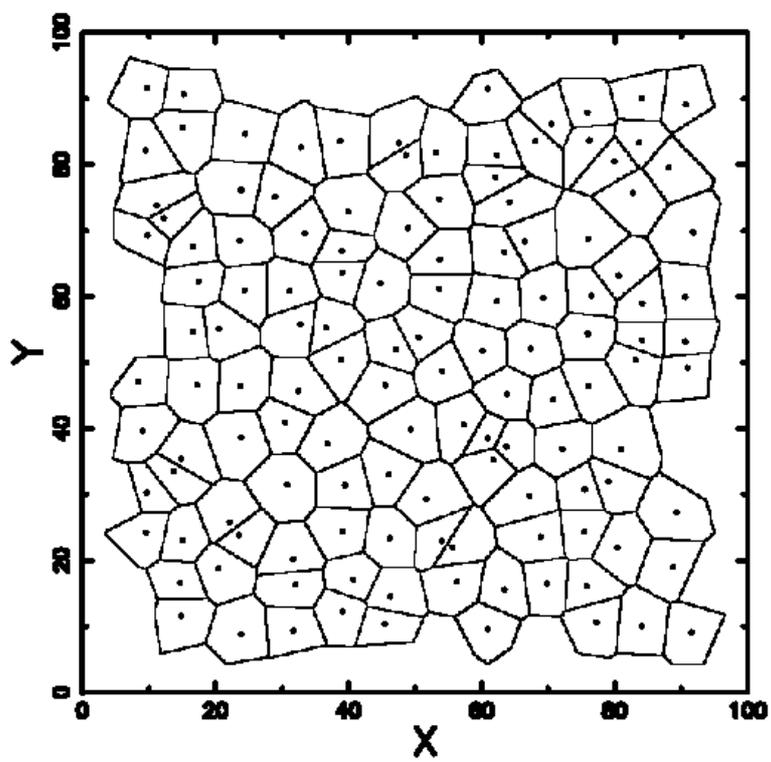}
\end {center}
\caption
{
An example of 2D tessellation generated
by  127  ACG seeds.
} \label{area2dlattice2v}
\end{figure}

\subsection{Applications of the adjustable seeds}

We present here  two applications of the adjustable seeds.
These adjustable  $3D$ seeds  can be calibrated
in order to have  $c=16$   for the Kiang distribution in volumes,
see PDF (\ref{kiang}).
To each value  of $s$ corresponds one value of $c$ that can be obtained from
the relation $c=1/\sigma^2$. For instance when  $s=0.316$, $c=32.82$ and
at when $s=1$, $c=15.87$; in particular
$c=16$ is obtained for  $s=0.64$.
The first application refers again to the local structure of
gas ($CO_2$) in the liquid-gas coexistence phase \citep{Idrissi2010}.
The parameter $c$ of the Kiang function for the PDF in volumes
can be parameterised as a
function of the  parameter $s$ as follows
\begin{equation}
c = C_2 s^{\alpha_2}
\quad  ,
\label{eqnalfa2}
\end{equation}
where $C_2$ and $\alpha_2$ can be found from
a simulation.
A numerical procedure gives  $C_2=24.39 $
and $\alpha_2= -0.44  $.
On equalizing the two equations (\ref{eqnalfa1}) and (\ref{eqnalfa2}) we obtain 
the following relationships between temperature, $T$, and
regulating parameter $s$
\begin{equation}
T = (\frac{C_2}{C_1})^{\frac{1}{\alpha_1}}
\, s^{\frac{\alpha_2}{\alpha_1}}
\quad ,
\label{tsrelationship}
\end{equation}
\begin{equation}
s = (\frac{C_1}{C_2})^{\frac{1}{\alpha_2}}
\, T^{\frac{\alpha_1}{\alpha_2}}
\label{strelationship}
\quad .
\end{equation}
The previous relationship allows to find the
theoretical standard deviation
of the Voronoi polyhedra volumes as function of the temperature.
We first fix the temperature as given
by the  values in Table \ref{tableidrissi} and
the relationship   (\ref{strelationship}) allows
to find $s$.
Given $s$ we obtain $c$  from   eqn.(\ref{eqnalfa2}) and
by the fact that for the Kiang's function
$\sigma^2=\frac{1}{c}$ we easily obtain $\sigma$
for a for the normalized variable.
The standard deviation for the non-normalized variable is
$\sigma_V=\langle V \rangle \sigma$.
Table \ref{tableadjustables} reports
the mean values and standard deviations of the
Voronoi polyhedra, $V$, as computed in \citep{Idrissi2010}
(first line) and  the procedure for the calculation of
standard deviations presented here
 (second line). In both lines $V$ are the values as given
in \citep{Idrissi2010} and are presented here to make
the comparison
easier. The third and fourth line present
the values of
$c$ of the Kiang function and the regulating parameter
$s$ of the ACG seeds.
Figure   \ref{acg_chemistry}
reports the chemical standard deviation as given by Chemistry
and the
theoretical standard deviation
as given by ACG as function of the temperature;
 it is clear that there is a good agreement between the calculated
$\sigma_V$ and the experimental standard deviation.
\begin{table}[ht!]
\caption
{
Mean values (first line),
standard deviations  (second line),
$c$ of the Kiang function (third line) and the regulating parameter
$s$ of the ACG seeds (fourth line).
}
\label{tableadjustables}
\begin{center}
\resizebox{12cm}{!}
{
\begin{tabular}{cccccccc}
\hline
 T/K& 250& 270& 285& 298& 303& 306& 313\\
\hline
$\langle V \rangle /{\AA}^3$~ Idrissi et al. &  69.7$\pm$  10.3&  77.5$\pm$  13.8&  87.1$\pm$
  19.4&  105.3$\pm$  31.1&  114.4$\pm$  38.2&
  156.8$\pm$  68.8&  156.8$\pm$  65.1\\
$\langle V \rangle /{\AA}^3$&  69.7$\pm$  9.19&  77.5$\pm$  15.032&  87.1
 $\pm$  22.142&  105.3$\pm$  33.463&  114.4$\pm$  39.51&
  156.8$\pm$  56.89&  156.8$\pm$  63.7\\
  c &  57.4&  26.5&  15.4&  9.9&  8.38&
  7.5&  6\\
 s &  0.143&  0.823&  2.808&  7.729&  11.275&
  14.101&  23.56\\
\hline
\end{tabular}
}
\end{center}
\end{table}
%
\begin{figure}
\begin{center}
\includegraphics[width=10cm]{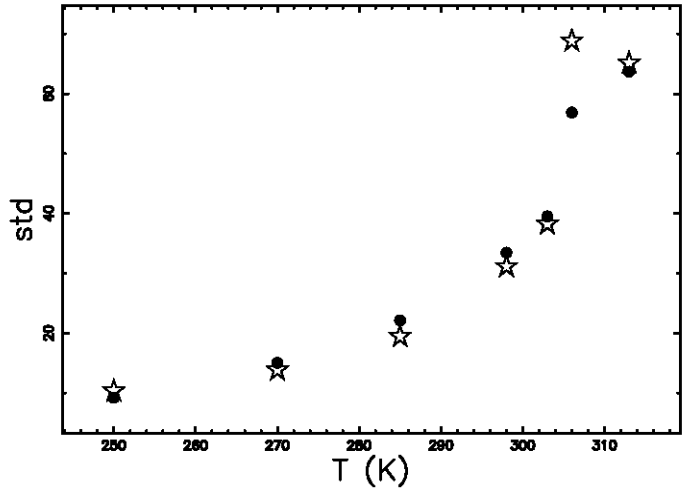}
\end {center}
\caption
{
Standard deviation as given
by Chemistry (empty stars) and
theoretical standard deviation
as given by variables volumes in ACG (full points)
 as function of the
temperature.
}
\label{acg_chemistry}
    \end{figure}
Next we consider the chords distribution of ACG tesselations,
and to this end  we adapt  the PDF  of Sobol's chords  as given
by (\ref{glbsobol}).
Calculated and empirical  distribution
functions of chords length
for ACG with
parameters as in Table \ref{parameters_seeds}
are shown in Figure
\ref{corda_df_lattice};
the K-S test gives $d_{max}=0.028$, $P_{KS}=0.0051$.
\begin{figure*}
\begin{center}
\includegraphics[width=10cm]{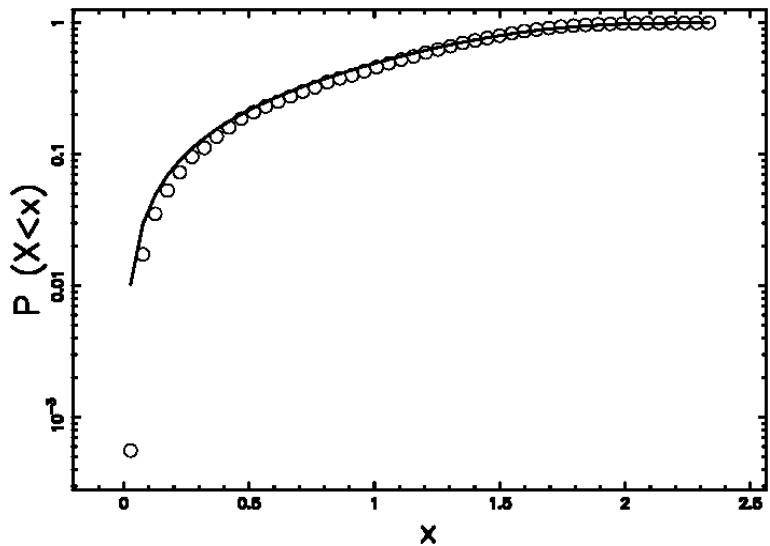}
\end {center}
\caption {
Data (empty circles) and theoretical
DF for  ACG seeds. The  theoretical
DF is the integral of  PDF ($g_{f,SVT}$)
(continuous line).
}
\label{corda_df_lattice}
    \end{figure*}

\section{Conclusion}

In this paper new types of three-dimensional  
Voronoi tessellations have been presented
whose centers are not sampled from an uniform distribution,
as in Poisson Voronoi (PVT) case,  but rather
are derived from  more regular sequences.
First  Sobol-Voronoi Tesselations (SVT) have been considered
in which  cells forming the partition have as
centers points generated by a Sobol
quasi-random sequence.
To make the notion of regularity of seeds configurations
more precise
a measure of uniformity has been presented.
In analogy with the case of PVT
we have used a generalized gamma distribution,
denoted with
$G_{SVT}$, to fit the volumes obtained by numerical simulations.
One should expect that the more regular configuration
of centers in the SVT
is reflected in the distributions
of cell volumes and this is indeed the case:
comparisons between the PDFs
show that
$G_{SVT}$  have smaller variance and are
more symmetric than the distribution  $G_{PVT}$
of volumes in respect to PVT.
It should be noted that volumes distributions  of PVT
cells can be fitted satisfactorily
by different types of gamma distributions, as mentioned
in Section \ref{pdfsec};
in contrast for volumes of SVT  only
the   generalized three-parameters  gamma provides a good fit.

As concerns applications,
SVT may be relevant in   modelling  partitions
of systems that
follow a more regular distribution than the usual
Poisson distribution and,
what is more interesting, transitions from ordered to
disordered states of the system
must be mirrored by  a corresponding  change from SVT to PVT.
An example has been presented in which volumes
occupied by molecules
of  $CO_2$ in  liquid-gas phases
undergo a transformation from regular to
more random distributions as the temperature increases.
Transitions from regular to disordered partitions of
space can be cast in a more  general setting
by considering the case in which center are first situated
on the nodes of a regular
 grid and then positions are perturbed with a 
gaussian noise regulated by a parameter
disorder $s$, thus creating an  Adjustable Cartesian Grid (ACG).
By increasing  $s$ one can reach first SVT distributions
and next PVT.
In this case a transition from ordered to disordered states
can be parametrized  by $s$.
Considering again to local structure of $CO_2$ a relation
has been derived between $s$ and $T$,
from which the standard deviation has been  computed
and next compared with the experimental one,
showing a good agreement.
Finally
statistics of chords resulting from intersections of
line with elements of PVT, SVT and ACG
have also  been investigated.
The interest of such type of statistics  resides in the fact that
in many experimental conditions only chords of
three-dimensional cells can be determined.
Results
show a good agreement between the analytical
formula, obtained with a  semi-empirical procedure and
data obtained from a simulation,
despite the approximations that have been used,
namely considering
cells to be a sphere and using an one-parameter
gamma distribution to
fit cells volumes,
from which chords distributions have been derived.

\section*{Appendix}

Numerical values of coefficients in Eq. (\ref{glbnonpoissonian}).

$$a_0={\pi }^{2/3}{8}^{2/3}\sqrt [3]{3}, \qquad a_1= 63997774118278000\,\Gamma  \left( 2/3 \right).$$

$$b_0=9161961861677625, \quad  b_1= 24431898297807000, \quad b_2=32575864397076000,$$

$$ b_3= 28956323908512000, \quad  b_4=19304215939008000, \quad b_5=10295581834137600, $$

k$$b_6=4575814148505600, \quad  b_7=1743167294668800, \quad b_8=581055764889600, $$

$$b_9=172164671078400, \quad b_{10}=45910578954240,
\quad b_{11}=11129837322240, $$

$$ b_{12}= 2473297182720, \quad b_{13}=507343011840,
\quad b_{14}=96636764160$$

$$b_{15}=17179869184.$$


\end{document}